\documentclass[
 10pt,
 reprint,
 floats,
 aps,
 amssymb,
 floatfix,
 onecolumn,
 nofootinbib,
 superscriptaddress]{revtex4-2}

\usepackage{amsmath,amsfonts,bm}

\newcommand{\ket}[1]{|#1\rangle}
\newcommand{\bra}[1]{\langle\,#1\,|} 
\newcommand{\bracket}[1]{\langle\,#1\,\rangle} 
\newcommand{\braket}[2]{\langle\,#1\, | \, #2\,\rangle}

\def\eqref#1{equation~\ref{#1}}

\def\1{\bm{1}}

\def\eps{{\epsilon}}

\DeclareMathAlphabet{\mathsfit}{\encodingdefault}{\sfdefault}{m}{sl}
\SetMathAlphabet{\mathsfit}{bold}{\encodingdefault}{\sfdefault}{bx}{n}

\newcommand{\E}{\mathbb{E}}

\usepackage[utf8]{inputenc}
\usepackage[english]{babel}

\newcommand{\cparagraph}[1]{\paragraph*{\textbf{\textup{#1}}}}

\usepackage[utf8]{inputenc} 
\usepackage[T1]{fontenc}    
\usepackage{hyperref}       
\usepackage{url}            
\usepackage{booktabs}       
\usepackage{amsfonts}       
\usepackage{nicefrac}       
\usepackage{microtype}      
\usepackage{xcolor}         

\usepackage{subfiles}
\usepackage[pdftex]{graphicx}
\usepackage{array}
\usepackage{multirow}

\begin{document}
\title{Learning ground states of quantum Hamiltonians with graph networks}

\author{Dmitrii Kochkov}
\affiliation{Google Research}
\author{Tobias Pfaff}
\affiliation{DeepMind}
\author{Alvaro Sanchez-Gonzalez}
\affiliation{DeepMind}
\author{Peter Battaglia}
\affiliation{DeepMind}
\author{Bryan K. Clark}
\affiliation{University of Illinois at Urbana-Champaign}

\begin{abstract}
  Solving for the lowest energy eigenstate of the many-body Schrodinger equation is a cornerstone problem that hinders understanding of a variety of quantum phenomena. The difficulty arises from the exponential nature of the Hilbert space which casts the governing equations as an eigenvalue problem of exponentially large, structured matrices. Variational methods approach this problem by searching for the best approximation within a lower-dimensional variational manifold. In this work we use graph neural networks to define a structured variational manifold and optimize its parameters to find high quality approximations of the lowest energy solutions on a diverse set of Heisenberg Hamiltonians. Using graph networks we learn distributed representations that by construction respect underlying physical symmetries of the problem and generalize to problems of larger size. Our approach achieves state-of-the-art results on a set of quantum many-body benchmark problems and works well on problems whose solutions are not positive-definite. The discussed techniques hold promise of being a useful tool for studying quantum many-body systems and providing insights into optimization and implicit modeling of exponentially-sized objects.
\end{abstract}

\maketitle

\section{Introduction}
Extreme eigenvectors play an important role in understanding the underlying structure of linear operators. In scientific computing the governing equations are often formulated as an eigenvalue problem of sparse, structured, high-dimensional matrices. In quantum physics such matrices represent Hamiltonians, which describe interactions between quantum degrees of freedom. Depending on the structure of the Hamiltonian matrix the underlying system might exhibit exotic properties such as superconductivity \cite{bardeen1957theory}, fractionalization \cite{laughlin1983anomalous} or topological ordering \cite{wen1990topological}. While writing down highly accurate governing equations is easy, finding even approximate solutions is extremely challenging due to the exponential scaling of the dimensionality of the underlying matrices.

%
While the final goal of studying the quantum many body problem is to develop effective analytical descriptions of the underlying processes, approximate numerical solutions are routinely used at all stages: from discovering model problems and providing structural insights to validating the results.

Variational Monte Carlo (VMC) is an important approach to tackling the quantum many body problem. In this approach the extreme eigenvectors of exponential size are parameterized implicitly using a compact representation. The goal is to find the best approximation to the exact answer within the selected variational manifold. This methodology has been critical in development of the theory of fractional quantum Hall effect \cite{laughlin1983anomalous}, superconductivity \cite{leblanc2015solutions}, etc.
The central challenge of the VMC approach is to design a suitable variational ansatz that can accurately capture the structure of the lowest energy eigenstate. This is commonly accomplished by deriving parameterized analytical solutions. Because constructing high quality analytical approximations is extremely challenging and introduces strong bias, there is a lot of interest in developing flexible parameterizations suitable for a wide class of problems.

Owning to the highly structured nature of the eigenvectors there has been a lot of interest in using neural networks to represent the lowest energy states. This is motivated by the hope that neural networks can take advantage of hidden structure of the eigenvectors and learn efficient, high-quality approximations. The other advantage of using neural networks is that they have lower bias compared to parameterized solutions due to their expansive expressivity; analytical solutions are often constrained to a particular physical phenomena.

At present, most of the neural network (NN) based wave-functions have been successful in two settings: (1) when the lowest energy state (ground state) is positive definite, and (2) when the NN-based model is combined with a standard variational ansatz that is well suited for approximating the phase structure of the target eigenstate. The former is restricted to the class of problems for which effective alternative methods exist \cite{reynolds1982fixed,sandvik1999stochastic}; the latter relies on the existence of high quality models and introduces a strong bias towards a particular structure. While progress has been made to learn approximations to the full complex-valued lowest energy states `from scratch'\cite{choo2019two,szabo2020neural,sharir2020deep}, current results have been largely limited to simple systems.

In this work we present a variational model based on graph networks that advances the state-of-the-art in learning lowest energy eigenstates on a diverse set of challenging benchmark problems among existing variational methods. Our key innovation is to learn distributed latent representations of wave-function amplitudes that respect the underlying physical symmetries of the problem. We demonstrate that this approach not only enables high quality approximations, but also makes it possible to extend the solutions to larger system sizes without retraining. This allows us to scale our model to approximate solutions of Heisenberg Hamiltonians with hundreds of quantum degrees of freedom while having lower variational energies than the best analytical constructions such as projected mean-field states.
\section{Background and related work}
In quantum mechanics a pure state of a physical system is described by a \emph{wave-function} $\ket{\psi}$ that assigns a complex-valued amplitude $\psi(c_{i})$ for each possible system configuration $c_{i}$. The set of distinct, simultaneously measurable system configurations $\left\lbrace c_{i} \right\rbrace$ is called a \emph{computational basis}, which is used to represent wave-functions as vectors in the Hilbert space $\ket{\psi}=\sum_{i}\psi(c_{i})\ket{c_{i}}$. All properties of the underlying state can be computed from the vector amplitudes $\psi(c_{i})\equiv\psi_{i}$; in particular $|\psi_{i}|^{2}$ corresponds to the unnormalized probability of observing the system in configuration $c_{i}$.

The size of the computational basis grows exponentially with the number of quantum degrees of freedom $N$. As a working illustration we consider an arrangement of $2$-state systems called spin-$\frac{1}{2}$. A single spin-$\frac{1}{2}$ can be observed in two states, up and down: $\lbrace \ket{\uparrow}, \ket{\downarrow} \rbrace$. System of two spins has four configurations: $\lbrace \ket{\uparrow \uparrow}, \ket{\uparrow \downarrow}, \ket{\downarrow \uparrow}, \ket{\downarrow \downarrow} \rbrace$. For $N$ degrees of freedom the size of the computational basis is $2^{N}$, which makes direct simulations of large systems prohibitively expensive.\footnote{Even storing the full vector describing the state of the system becomes impossible for large $N$}

\begin{figure}
  \centering
  \includegraphics[width=1.0\linewidth]{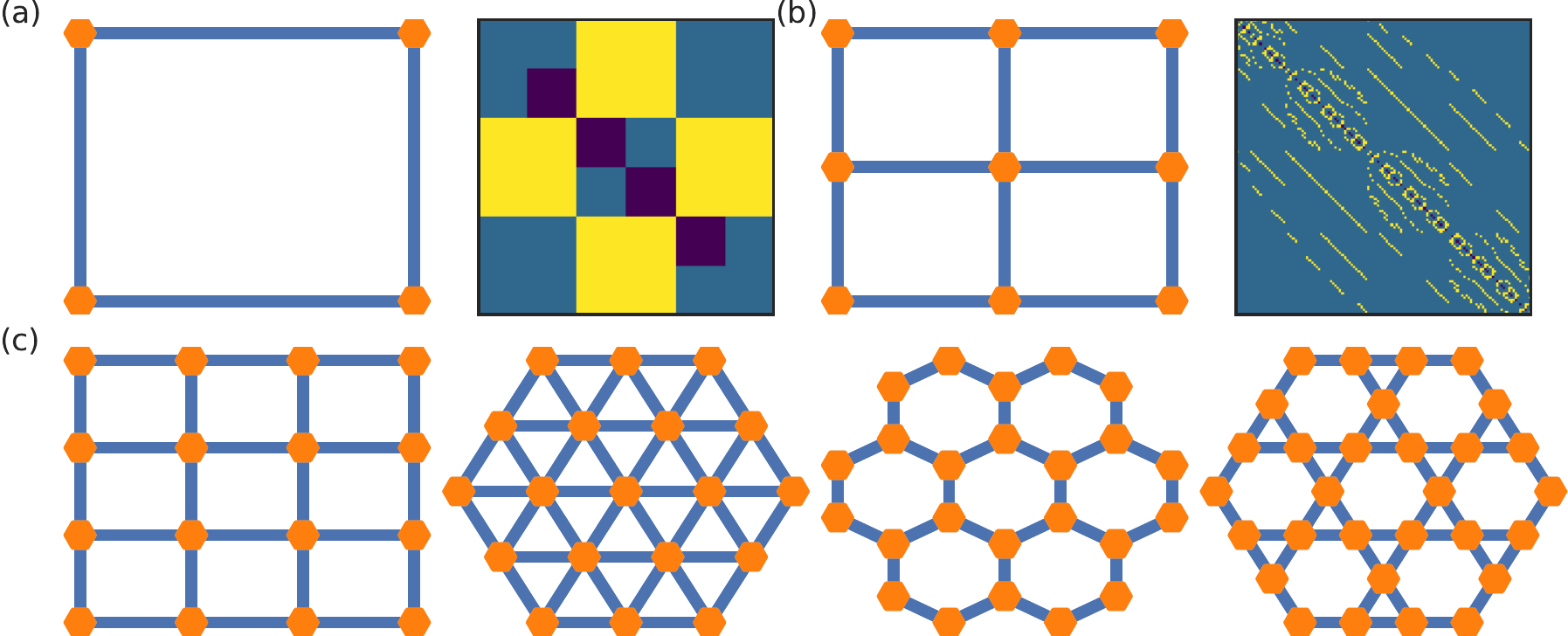}
  \caption{(a)-(b) Visualization of small unit cells of the square lattice and associated Heisenberg Hamiltonian matrices, whose dimensionality grows exponentially with the system size $N$ as $C(N, N//2)$.
  (c) Visualization of square, triangular, honeycomb and kagome lattices. Orange dots denote spin-$\frac{1}{2}$ degrees of freedom and blue lines indicate exchange interactions.}\label{fig:lattices}
\end{figure}

In equilibrium, quantum systems settle into distinct phases, determined by the interactions between the underlying degrees of freedom. The interactions are described by a Hamiltonian operator $\hat{H}$, or equivalently the Hamiltonian matrix $H_{ij}=\bra{c_{i}} \hat{H} \ket{c_{j}}$. Because interactions are typically restricted to spatially nearest neighbors, the Hamiltonian matrix is usually sparse and highly structured. The structure is determined by the spatial patterns in which the degrees of freedom are arranged and the nature of the interactions. Spatial lattices considered in this work, together with sample small scale Hamiltonian matrices are shown in Fig. \ref{fig:lattices}.

The $H_{ij}$ matrix defines the relationship between energy levels $E_{n}$ and energy eigenstates $\ket{\psi_{n}}$ via the stationary Schrodinger equation (Eq. \ref{eq:schrodinger}). In the rest of the paper we use the Einstein notation in which repeated indices are assumed to be summed over: $\bra{c_{i}} \hat{H} \ket{\psi} = \sum_{j} H_{ij} \psi(c_{j}) \equiv H_{ij} \psi_{j}$.
\begin{equation}
    \hat{H}\ket{\psi_{n}}=E_{n}\ket{\psi_{n}}
    \Longleftrightarrow
    \sum\limits_{j} H_{ij}\psi_{n}(c_{j}) = E_{n} \psi_{n}(c_{i}) \label{eq:schrodinger} 
\end{equation}
The equilibrium properties of the system are dominated by the low energy eigenstates. In many cases the nature of the underlying phase of matter can be understood from the hidden patterns encoded in the lowest energy eigenstate $\ket{\psi_{0}}$.

\subsection{Variational Monte Carlo}\label{subsec:vmc}
Variational Monte Carlo (VMC) is an efficient technique for approximating lowest energy eigenstates. It addresses the exponential complexity of the Hilbert space $\mathcal{H}$ by:
\begin{enumerate}
    \item Working with low dimensional variational manifold $\mathcal{V} \subseteq \mathcal{H}$
    \item Using stochastic sample-based approximations for evaluation and optimization
\end{enumerate}
The manifold $\mathcal{V}$ is defined implicitly (Eq. \ref{eq:variational_ansatz}) by a set of vectors $\ket{\psi_{f, w}}$, where $f$ is a variational model (commonly referred to as ansatz) and $w$ are the variational parameters. 
\begin{equation}
    \ket{\psi_{f, w}} = \sum_{i}f(w, c_{i})\ket{c_{i}} \qquad f: w, c_{i} \rightarrow \psi_{f, w}(c_{i}) \qquad \mathcal{V} = \lbrace \ket{\psi_{f, w}}, \ w \in \mathbb{R}^{m} \rbrace \label{eq:variational_ansatz}
\end{equation}
While specifying a wave-function using a variational ansatz is easy, exact evaluation of its properties still requires an exponential number of queries to the function $f$. This complexity is alleviated by replacing exact evaluations with stochastic approximations based on Monte Carlo sampling.

The goal of the VMC algorithm is then to find the best parameters $w^{*}$ that correspond to vector $\ket{\psi_{f, w^{*}}}$ that most closely approximates the lowest energy eigenstate $\ket{\psi_{0}}$.
The success of VMC approach heavily depends on the quality of the variational ansatz $f$.

\cparagraph{Variational models} Variational ansatz $f$ are designed to be: (1) \emph{expressive} and (2) \emph{easy to optimize}. The expressiveness enables accurate approximations, as well as reduces the bias introduced by the model. Optimizability makes it possible to efficiently navigate the variational manifold in the search for the best approximation of $\ket{\psi_{0}}$. This motivates the use of neural networks as variational ansatz - they are highly expressive and amendable to optimization with gradient based methods.

\cparagraph{Evaluation} VMC uses stochastic estimates of physical observables to evaluate the performance. Ideally, the quality of the approximation $\psi_{f,w^{*}}$ would be measured by estimating the overlap $\mathcal{O}=|\braket{\psi_{0}}{\psi_{f,w^{*}}}|$ between the normalized variational state $\ket{\psi_{f,w^{*}}}$ and the exact lowest energy eigenvector $\ket{\psi_{0}}$. Since $\ket{\psi_{0}}$ is not known, the success is typically evaluated by the Rayleigh quotient $\bracket{E}$ (Eq. \ref{eq:energy}), commonly referred to as energy expectation.
\begin{equation}\label{eq:energy}
    \bracket{E} = \frac{\bra{\psi} \hat{H} \ket{\psi}}{\braket{\psi}{\psi}} = \frac{\sum_{i}|\psi_{i}|^{2} (\left[H_{ij}\psi_{j}\right] / \psi_{i})}{\sum_{i}|\psi_{i}|^{2}} = \E_{|\psi|^{2}}(\left[H_{ij}\psi_{j}\right]) / \psi_{i}
\end{equation}
The lowest energy eigenstate $\ket{\psi_{0}}$ establishes the lower bound $\bracket{E}=E_{0}$ based on the Schrodinger equation (Eq. \ref{eq:schrodinger}), hence states with lower values of $\bracket{E}$ are considered better approximations.

Note that the final expression for $\bracket{E}$ consists of the ratio of terms weighted by the probability distribution $|\psi|^{2}$. This enables efficient estimation using important sampling techniques such as Markov-Chain Monte Carlo. The process boils down to computing the average of $\left[H_{ij}\psi_{j}\right] / \psi_{i}$ for system configurations $c_{i}$ sampled from the probability distribution $|\psi|^{2}$. We will use the notation $\E_{|\psi|^{2}}$ to refer to such averaging. While in principle the matrix-vector product $\left[H_{ij}\psi_{j}\right]$ could contain exponentially many terms, real world Hamiltonians are sparse and it is tractable to compute all non-zero entries.

\cparagraph{Optimization} The variational parameters $w$ can be optimized by several methods. The two most common techniques are based on (1) energy gradient and (2) imaginary time evolution. In energy gradient the parameter update $\delta w$ is driven by the gradient of the stochastic estimate $\nabla_{w}\bracket{E}$, optimizing the evaluation metric. In imaginary time evolution $\delta w$ is chosen such, as to maximize the overlap between $\ket{\psi_{f,w + \delta w}}$ and $(\bm{1}-\beta H_{ij})\ket{\psi_{f,w}}$. This procedure is motivated by the fact that $\ket{\psi_{0}}$ is a fixed point of the matrix exponential operator $\exp(-\beta( H_{ij} - E_{0})) \ as \ \beta \rightarrow \infty$.

We note that in the VMC setting operations of computing the gradient $\nabla_{w}\left[\Box\right]$ and the expectation value $\bracket{
\Box}$ do not commute, as the gradient must take into account how the change of model parameter $w$ affects the probability distribution $|\psi_{f, w}(c_{i})|^{2}$. These details, together with an overview of the two optimization methods are discussed in Appendix \ref{apx:optimization_methods}.
\subsection{Related work}
There has been a large body of research focused on improving and analyzing properties of the quantum states represented with neural networks \cite{carleo2017solving,kochkov2018variational,choo2019two,nomura2020dirac,nomura2017restricted,sharir2020deep,ferrari2019neural,hibat2020recurrent}. The two most active research threads include (1) extending the applicability of fully learned wave-functions  \cite{kochkov2018variational,szabo2020neural,choo2019two,hibat2020recurrent,valenti2021correlation,yang2020scalable} which learn the wave-function without the input of a-priori physics knowledge and are therefore unbiased and (2) a hybrid model using neural networks in combination with standard variational models which incorporate assumptions about the underlying physics by starting with a simplified wave-function which is in the correct phase \cite{luo2019backflow,nomura2017restricted,nomura2020dirac,ferrari2019neural}. Our work primarily contributes to the former goal developing wave-functions which are not biased by a-priori assumptions about the physics, while performing on par or better with respect to most hybrid models.

To enable learning of both magnitudes and phases a common approach is to use complex-valued network weights and custom activation functions \cite{choo2019two,choo2018symmetries,valenti2021correlation,deng2017machine}. While for many cases exact solutions were analytically constructed, many works found the direct optimization difficult, resulting in the need to initialize the model with a particular sign structure. A different approach introduced in parallel with this work is based on using separate neural networks to predict the magnitude and the phase of the wave-function \cite{szabo2020neural}, where authors compute the phase of the wave-function as the argument of a sum of phasors. Our approach is different, as we predict phase of the wave-function directly. Our experiments show that this choice is critical to enable effective generalization of the learned sign structure.

Similar to \cite{kochkov2018variational,yang2020scalable} our models use sum-based reduction to enable system size generalization. Here we show that this technique effectively applies not only to scaling the magnitudes, but also the phase information. This suggests that when empowered with proper structural mechanisms, neural networks can generalize when predicting the sign structure, which has been an active discussion topic \cite{westerhout2020generalization}.

While in this paper we test our models in the context of quantum spin systems, there has been a substantial effort showing that similar approaches could be applied to problems of quantum chemistry. For example, in \cite{choo2020fermionic} authors show how molecular problems can be approached by mapping fermionic problems onto spin Hamiltonians. Other examples of applying ML-powered variational ansatz to quantum chemistry include PauliNet \cite{hermann2020deep}, FermiNet\cite{pfau2020ab}.

\section{Model}\label{sec:model}
\begin{figure*}
  \centering
\includegraphics[width=1.0\linewidth]{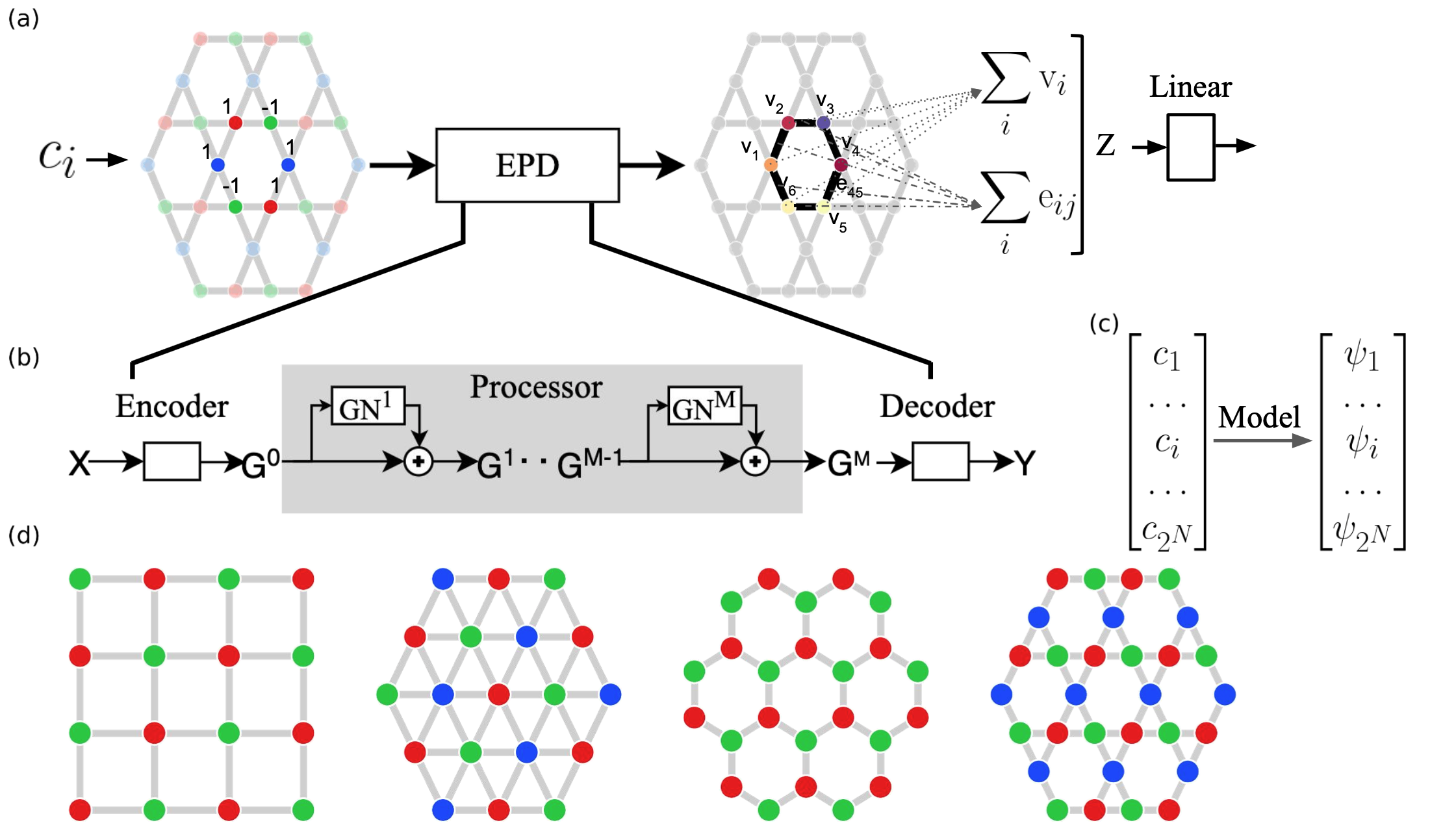}
  \caption{(a) Our model predicts scalar outputs ($\log/arg(\log(\psi_{i}))$) from an input basis element $c_{i}$ represented as a graph. (b) \emph{Encode-process-decode} module maps sublattice augmented input configuration $X$ to a distributed latent representation $Y$. (c) Mapping from basis elements to amplitudes implicitly defines a full vector. (d) Example of sublattice encodings $s$ used in this work as auxiliary inputs.
  }\label{fig:network}
\end{figure*}

We specify a variational model $f$ introduced in Section \ref{subsec:vmc} using a deep neural network. The network representing $f$ implicitly defines a complex-valued vector
$\ket{\psi_{f,w}}=\sum_{i}\psi_{f,w}(c_{i})\ket{c_{i}}$, by establishing a mapping between basis configurations $c_{i}$ and corresponding amplitudes $\psi_{f,w}(c_{i}) \in \mathbb{C}$.

For a systems of $N$ spin-$\frac{1}{2}$ degrees of freedom, each $c_{i}$ is represented by $N$ values in $\lbrace \pm1 \rbrace$, indicating the configuration of each spin. Our model computes $\psi_{f,w}(c_{i})$ by predicting the scalar values $\log(|\psi_{f,w}(c_{i})|)$ and $\arg(\psi_{f,w}(c_{i}))$ as outputs. The choice to work with the \emph{logarithm} of the wave-function rather than $\psi(c_{i})$ directly is motivated by the same numerical stability considerations as in using log-likelihood in generative models. This is equivalent to choosing $\exp$ as final activation function, motivation for which is discussed in Appendix \ref{apx:physical_priors}.

\cparagraph{Architecture} Our model uses graph neural networks (GNN) to compute $\log(|\psi_{f,w}(c)|)$ and $\arg(\psi_{f,w}(c))$ from an input basis configuration $c \in \lbrace c_{i} \rbrace$ in four steps, also shown in Fig.\ref{fig:network}(a)
\begin{enumerate}
    \item Input basis element $c$ and a static sublattice encoding $s$ are used to initialize a graph $X$
    \item Distributed latent representation $Y$ is computed via an \emph{encode}-\emph{process}-\emph{decode} graph network
    \item Final latent vector $Z$ is obtained by concatenating sum-pooled node and edge features
    \item Latent vector $Z$  is processed with a linear layer to predict scalar outputs
\end{enumerate}
First, an initial graph tuple $X=(v_{i}, e_{jk})$ is constructed based on the input configuration $c$ and fixed sublattice encoding $s$. Nodes $v_{i}$ are initialized by concatenating $c$ and $s$, and edges $e_{jk}$ between interacting spins are initialized to zero. The sublattice encoding $s$ is a one-hot vector, denoting the node's location in a unit cell that respect all of the symmetries of the lattice up to a relabeling of the one-hot vectors. This concept is related to the algebraic symmetry group of the lattice and has been successfully used in constructing variational states for classical \cite{messio2011lattice} and quantum\cite{wen2002quantum} systems. We show some of the sublattices used in this work in Fig. \ref{fig:network}(c) and discuss it in more detail in Appendix\ref{apx:sublattice_encodings}.

Next, we use a GNN model with \emph{encode-process-decode} \cite{battaglia2018relational} architecture (Fig \ref{fig:network}(b)) to compute a distributed latent representation $Y$. The \textbf{encoder} maps nodes and edges of input graph $X$ to latent features of width $64$. Next, the \textbf{processor} performs $M$ rounds of message passing, updating the graph as $G_{t} = G_{t-1} + GN_{t}(G_{t-1})$. The residual GraphNet blocks $GN_{t}$ \cite{sanchez2018graph} use MLP node and edge functions. The \textbf{decoder} computes the final latent edge and node features $Y$ from $G_M$, by processing them using a MLP. More details on the architecture can be found in Appendix \ref{apx:model_details}.

Finally, we obtain scalar outputs by sum-pooling the latent features over nodes and edges; concatenating the results into a single vector $Z$ of size $128$; and processing it with a linear layer. We consider two model variants: (1) GNN, an ansatz that uses a single network with the final linear layer of size two; (2) GNN-2, an ansatz which computes each output using a separate network with output size of one. Other variations are discussed in  Appendix \ref{apx:model_variations}.

Graph neural networks are a natural choice for modeling this problem, as they implicitly embed the spatial symmetries of the lattice into the variational model. Combining GNNs with sum-based reduction makes our model `size consistent' -- the variational manifold automatically factorizes into a product of individual states if the physical system is separated into non-interacting components. Exploiting the underlying structure of the problem not only makes the problem easier to solve, but improves generalization, and even allows zero-shot transfer to problems of larger size.

\cparagraph{Evaluation} Because our model has the same input/output structure as traditional variational ansatz, we can embed them inside of a VMC code, which consists of (1) sampling configurations $c_{i}$ from the probability distribution $\propto|\psi_{f,w}(c_{i})|^{2}$ and (2) computing physical quantities such as $H_{ij} \psi_{j}$. The sampling is accomplished via Metropolis-Hastings algorithm, where new configurations are accepted or rejected based on the detailed balance condition. The observables such as $H_{ij} \psi_{j}$ can be computed directly, since Hamiltonian matrices are sparse. To obtain accurate estimates we run multiple sampling chains in parallel. All MCMC hyperparameters are summarized in Appendix \ref{apx:mcmc}.

\cparagraph{Training} We optimized the parameters of our models using imaginary time supervised wave-function optimization (IT-SWO)\cite{kochkov2018variational}. This approach works iteratively, optimizing the parameters $w$ of the model to maximize the overlap between the current variational ansatz $\ket{\psi_{f,w}}$ and an imaginary-time evolved state $\ket{\phi}=(\bm{1}-\beta \hat{H})\ket{\psi_{f,r}}$ with frozen parameters $r$. Each iteration is accomplished via an inner loop of multiple steps. Depending on the length of the inner loop, IT-SWO interpolates between following the energy gradient $\nabla_{w}\bracket{E}$ and the natural gradient\cite{amari1998natural} $\tilde{\nabla}_{w}\bracket{E}$. We used an intermediate value of $30$ updates per iteration which worked well for models with large number of parameters. More details on the optimization are discussed in Appendix \ref{apx:optimization}.

\section{Test problems} \label{sec:test_problems}
To evaluate the performance of our models we consider a set of Heisenberg Hamiltonians (Eq. \ref{eq:heisenberg_hamiltonian}), where $\langle i,j \rangle$ and $\langle \langle i,j \rangle \rangle $ corresponds to indices of nearest and next-nearest neighbors respectively.
\begin{equation}
    \label{eq:heisenberg_hamiltonian}
    \hat{H} = \sum\limits_{\langle i,j \rangle} \hat{S}_{i} \hat{S}_{j} + J_{2} \sum\limits_{\langle \langle i,j \rangle \rangle} \hat{S}_{i} \hat{S}_{j} 
\end{equation}
The lowest energy states of such Hamiltonians are strongly influenced by the underlying geometry of the problem and relative interaction strength $J_{2}$ between next nearest and nearest neighbors. Our test suite includes four routinely studied geometric variations: \emph{square}\cite{richter1994violation,nomura2020dirac,gong2014plaquette,hu2013direct,choo2019two}, \emph{honeycomb}\cite{mosadeq2011plaquette,li2012phase,albuquerque2011phase}, \emph{triangular}\cite{iqbal2016spin,kaneko2014gapless,wietek2017chiral,wu2019randomness} and \emph{kagome}\cite{hu2015variational,kolley2015phase,changlani2018macroscopically} lattices. All of these models feature complex patterns in the lowest energy spectrum for which a lot of analytical and numerical methods have been developed.

Each of these Hamiltonians features multiple phases, including disordered domains. For each phase a different structure emerges in the lowest energy eigenstates. For \emph{square} and \emph{honeycomb} lattices the disorder regime occurs at $J_{2}=0.5$ and $J_{2}=0.2$, induced directly by interactions with next nearest neighbors. \emph{Triangular} and \emph{kagome} lattices additionally feature geometric frustration, leading to even larger quantum fluctuations and harder to model structures. We provide a more detailed account for these problems in Appendix \ref{apx:test_problems_details}.

\begin{figure}
  \centering
  \includegraphics[width=1.0\linewidth]{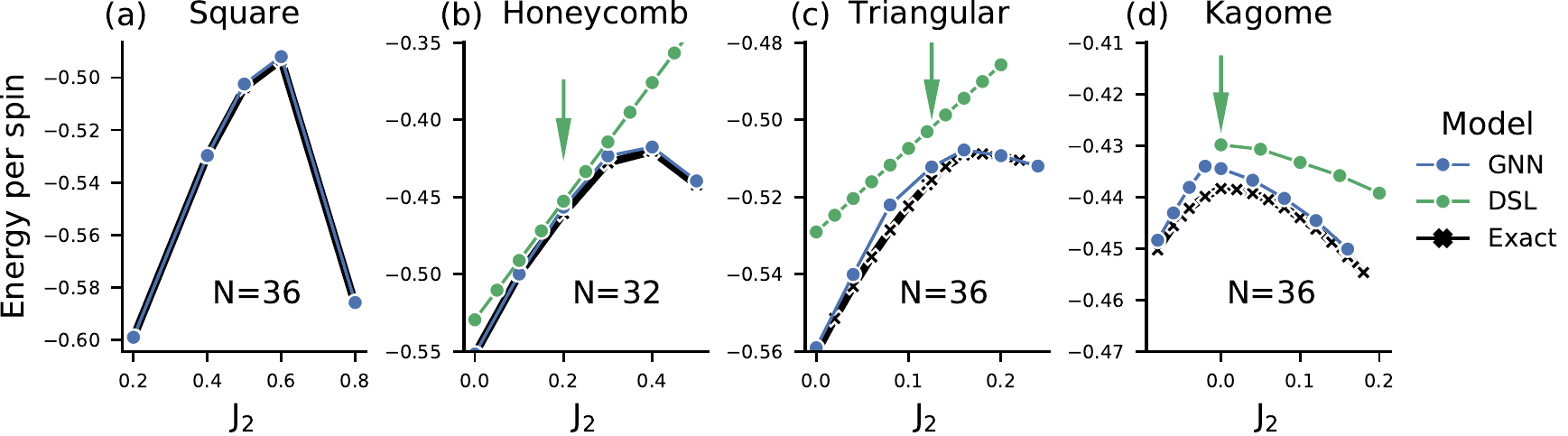}
  \caption{Our models accurately approximate the structure of the lowest energy state across different problems and interaction parameters. Columns show energy per spin as a function of relative interaction strength $J_{2}$ between next and nearest neighbors on different lattices. System sizes $N$ are annotated on each plot. Color differentiates exact solutions, our GNN models and analytically constructed DSL model, provided for a reference. Arrow indicates the value of $J_{2}$ for which DSL model is believed to be most appropriate. GNN-2 model (not shown) achieves similar results to GNN on this evaluation.}\label{fig:model_performance_36}
\end{figure}
\section{Results}\label{sec:results}
We tested the performance of our models on the test suite of frequently studied Heisenberg Hamiltonians (Section \ref{sec:test_problems}). For each Hamiltonian specification we trained our models and compared their performance to ML-based and traditional variational models from the literature. Our main finding is that our models are able to approximate lowest energy states with high accuracy for a variety of different problems without relying on any prior knowledge about the structure of the target state, as well as scale and even generalize to larger system sizes.

We first evaluated the performance of our models on the test problems with system size of $36/32$. This is close to the largest system size for which exact solutions are feasible. We found that our models approximate the lowest energy states with high accuracy, both across different systems and interactions strengths. In Fig. \ref{fig:model_performance_36} we show variational energies of our GNN model compared to the exact energies and best known analytical baselines developed for the disordered regimes. The GNN-2 variant produces identical results on this task and is omitted from the plot. The close tracking of energy as we change the interaction strength $J_{2}$ reassures that the model is flexible enough to discover the transitions between different hidden structures. To the best of our knowledge this is the first time a single NN-based model is shown to accurately approximate solutions across different interaction parameters on a non-square geometry without building on top of an existing ansatz.

\begin{figure}
  \centering
  \includegraphics[width=1.0\linewidth]{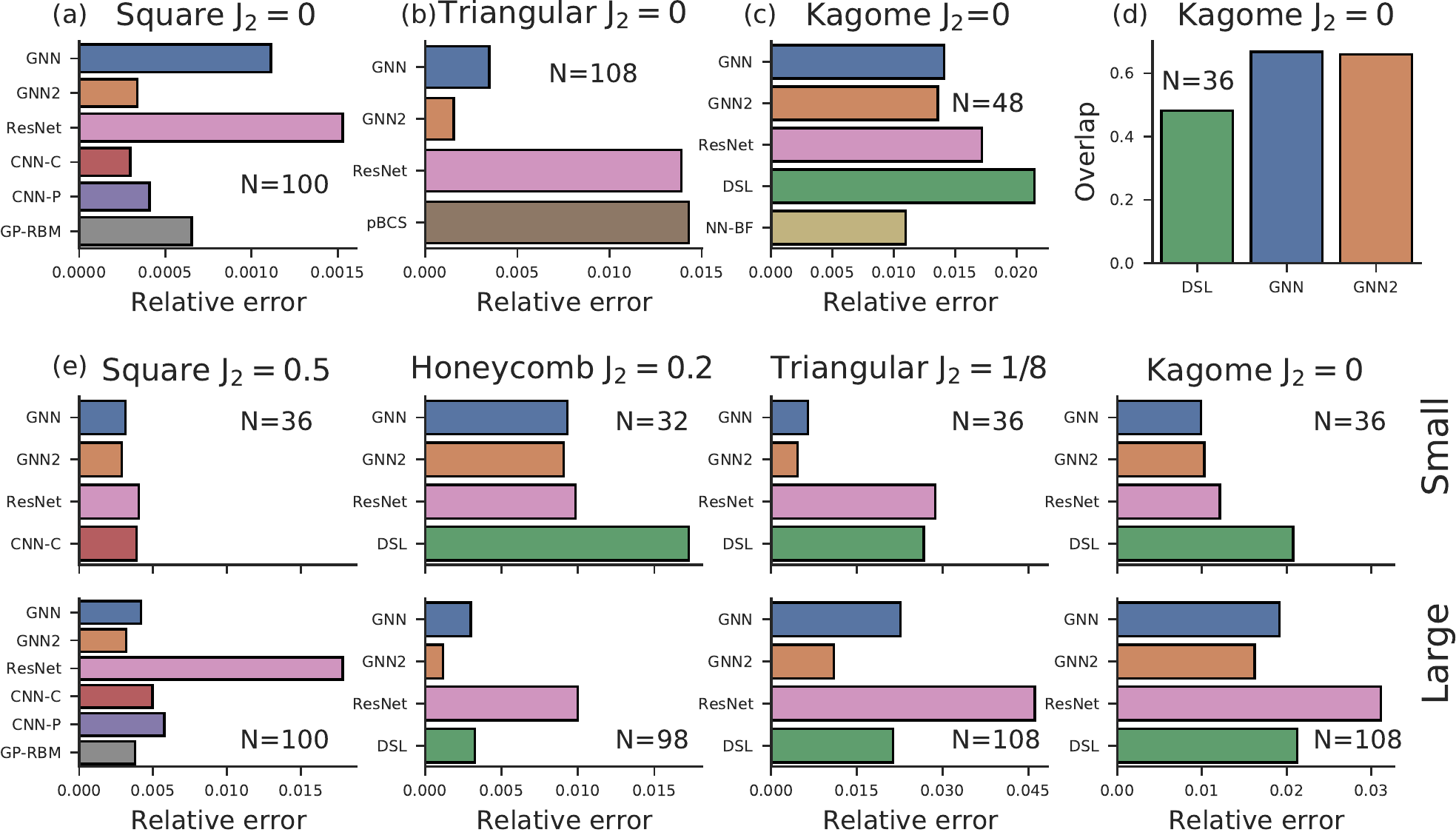}
  \caption{Our models compares favorably to learned and analytical baselines on a wide range of test problems. (a)-(b) Relative error of variational energies with respect to the estimated ground state energies of our model and baselines in ordered regime on \emph{square} and \emph{triangular} lattices. (c) Comparison of variational energies on \emph{kagome} lattice with $48$ spins. (d) Comparison of the overlap with the exact lowest energy eigenstate to a DSL ansatz on a 36-site \emph{kagome} lattice. (e) Relative error of variational energies in the harder-to-model disordered regimes for \emph{square}, \emph{honeycomb}, \emph{triangular} and \emph{kagome} lattices on small and large systems. The system sizes $N$ are annotated on each plot.}\label{fig:model_comparison}
\end{figure}

\cparagraph{Model comparison} We compared our models to a suite of analytical variational models: Dirac spin liquid (DSL)\cite{iqbal2013gapless}, projected BCS (pBCS)\cite{kaneko2014gapless}; pure ML-based wave-functions: (CNN-C)\cite{choo2019two}, (CNN-P)\cite{szabo2020neural}; hybrid approaches: neural Gutzwiller-projected ansatz (GW-RBM)\cite{ferrari2019neural}, neural network backflow (NN-BF)\cite{luo2019backflow}; as well as our own baseline ResNet, which replaces the graph-network part of our model with a MLP-based residual neural network. We use relative error $\eps = \frac{E - E_{0}}{E_{0}}$ as the comparison metric, where for small system sizes $E_{0}$ denotes the exact ground state energy and $E_{0}^{*}$ denotes the extrapolated values for large system sizes (specified in Appendix \ref{apx:ground_state_energies}). Fig. \ref{fig:model_comparison} (a)-(b) shows the performance on \emph{square} and \emph{triangular} lattices at $J_{2}=0$, when the system is in the ordered phases. We find that both GNN and GNN-2 models perform well on this task (relative errors $~10^{-3}$), with GNN-2 ansatz consistently fine tuning closer to the ground state energy.

Next, in Fig.\ref{fig:model_comparison} panel (e) we compare model performance at Hamiltonian configurations that feature disordered phases, which are much harder to model. We find that our models achieve similar performance to other ML-based approaches on the \emph{square} lattice\footnote{The best model for the square lattice is RBM-PP \cite{nomura2020dirac}, which we use as an extrapolated ground state energy.}, while also being able to efficiently tackle other geometries. In particular, our model achieves almost $50$\% higher overlap with the ground state on the \emph{kagome} lattice on $36$-site system compared to DSL (Fig.\ref{fig:model_comparison}(d)) and almost matches the performance of a hybrid approach NN-BF\cite{luo2019backflow} (Fig.\ref{fig:model_comparison} (c)) on $48$ spin system. As the system size is increased, we find that in the disordered regime the performance gap reduces, which suggests that additional mechanisms might be needed to effectively capture the underlying structures.

When compared to the ResNet baseline, our model performs similarly on small system sizes, but features a significant improvement on larger clusters. We speculate that this is related to the ability of the distributed representation to uniformly take into account contributions from all parts of the system, while an unstructured model needs to learn those explicitly.

\begin{figure}
  \centering
  \includegraphics[width=1.0\linewidth]{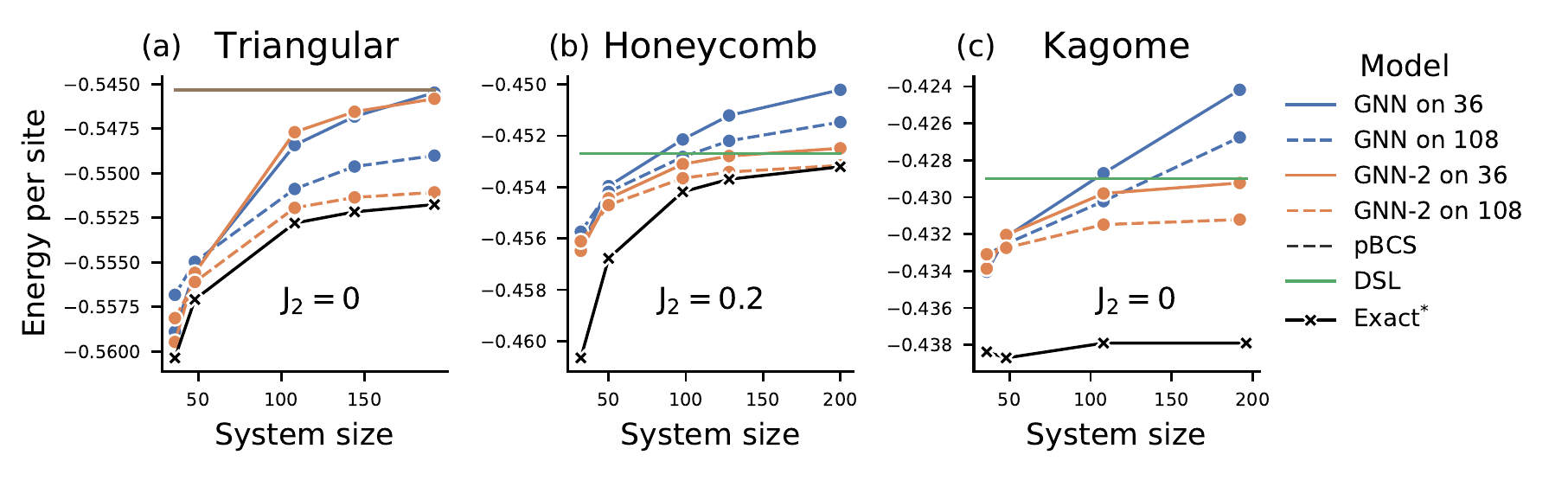}
  \caption{By learning distributed representations our models can generalize across system sizes without significant loss of accuracy. Columns show variational energies per spin as a function of system size for models with frozen weights on different lattices. Line style differentiates the system size on which models were trained on. We also provide extrapolated exact energies and traditional variational models for reference. The relative interaction strength $J_{2}$ is annotated on each plot. Error bars are smaller than the marker size.}\label{fig:system_size_generalization}
\end{figure}

\cparagraph{System size generalization}
Because the computation learned by graph networks is local, our models can be applied to systems of different size without retraining. When variational wave-function is evaluated a on a larger system, it needs to accurately capture an exponentially larger space of amplitudes to have good variational energies. This requires accurate predictions of both magnitudes and highly oscillatory phases for each configuration in computational basis. Our models are able to accomplish this because they predict both components that are `size consistent' by construction and combine contributions from each subsystem uniformly. In Fig. \ref{fig:system_size_generalization} we show the performance of our models across system sizes. We note that the system size on which the models were optimized affects the generalization, with models optimized on smaller system sizes performing slightly worse on larger systems and vice-versa. This is to be expected, as to achieve the lowest energy state on a small system the model needs to exploit the finite-size effects of the physical system, which are not system size invariant. The fact that the rules learned from a system of a different size generalize to larger domains further suggests that GNN-based models discover the underlying patterns. This property of the model is important as it allows to greatly simplify training on large system sizes by starting with an already good approximation. In our evaluations we found that GNN-2 model is consistently better at obtaining lower energies on larger domains and seems to be better at generalizing to larger system. We suggest that this is related to the coupling of variational parameters responsible for highly oscillatory phases and slower varying magnitudes in the GNN model.

\begin{figure}
  \centering
  \includegraphics[width=1.\linewidth]{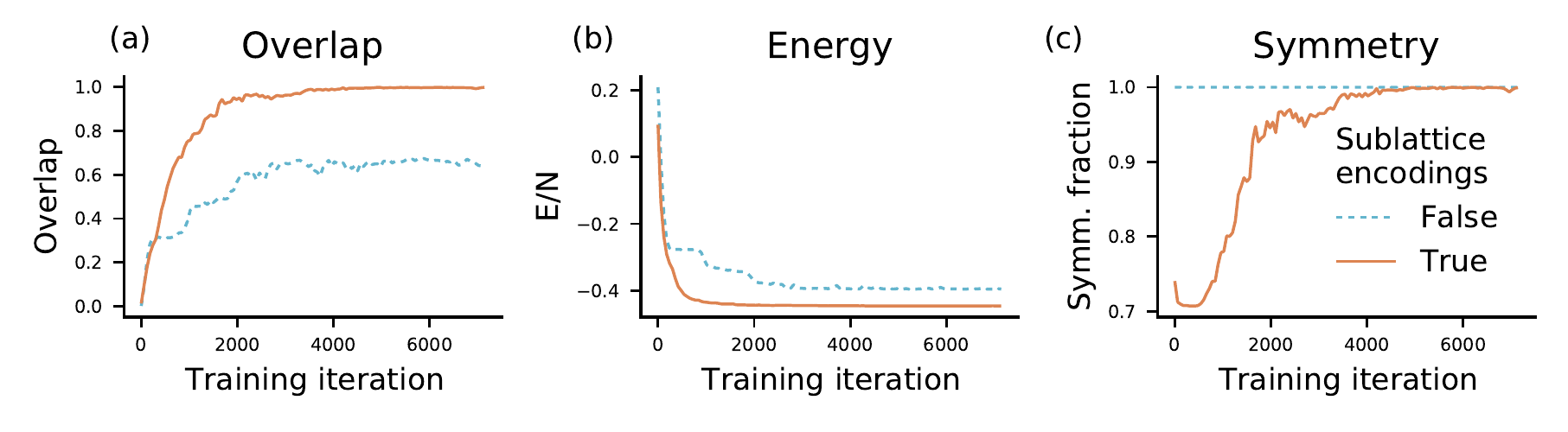}
  \caption{Including sublattice encodings dramatically improves the performance of the GNN models. (a) Overlap between the exact lowest energy state and the model as a function of training iterations. (b) Energy throughout the optimization process. (c) Fraction of the symmetric component throughout the optimization. Color indicates whether sublattice encodings are used. }\label{fig:sublattice_ablation}
\end{figure}

\cparagraph{Sublattice encoding ablation}
We found that including sublattice encodings to the input of a graph network is critical to achieve high approximation accuracy and enable generalization across system sizes. Naively, it would seem that the role of providing the sublattice codes is to simply break the symmetry of the solution. We note that this is not the case. For the ablation study we consider a simple $1$D lattice with $16$ degrees of freedom. The true solution is symmetric with respect to all of the lattice transformations that leave the Hamiltonian invariant. We train two models that differ only by the presence of the sublattice encodings. We inspect the symmetry properties of the models throughout the optimization, their performance measured by overlap with the exact lowest energy eigenstate $\mathcal{O}=\braket{\psi_{0}}{\psi_{nn}}$, as well as energy expectation $\langle E \rangle$. The result are presented in Fig. \ref{fig:sublattice_ablation}. The model that uses sublattice encodings performs significantly better at approximating the target state. Note that towards the end of training the model restores the symmetry broken by the sublattice encodings, as indicated in panel (c) of Fig. \ref{fig:sublattice_ablation}. A fully symmetric model, despite being naturally placed in the correct symmetry sector, struggles to find an accurate approximation. Interestingly, this problem does not arise in a simpler setting when the model is provided with exact phases (sign structure), suggesting that it is the modeling of the highly oscillatory phases that benefits from being able to represent the underlying structure in the symmetry-broken hypothesis space and "restore" the symmetry. In Appendix \ref{apx:sublattice_encodings} we provide additional details and discuss sublattice encodings in the light of non-universality of graph networks.

\section{Conclusions}
We presented a simple - yet efficient - machine learning framework for learning high quality approximations of lowest energy solutions of the quantum many body problem. Our model uses graph networks to learn a distributed representation of the wave-function that respects the underlying symmetries of the problem. Our experimental results demonstrate that the same architecture is able to accurately capture the changes in the ground states of Heisenberg models across different lattices and interaction patterns, advancing the state of the art results on several challenging problems. Of particular interest is the ability to accurately represent ground state wave-functions which have complicated sign structures. A key strength of our approach is that it does not rely on a-priori knowledge of physics avoiding the bias that is introduced in using such knowledge.  We find that learning distributed representation not only helps achieve higher accuracy, but also enable generalization to larger problem sizes without significant loss of accuracy.

It is worth noting that the presented approach still features limitations related to variational nature of the method. We believe that the proposed models are most suited for studying quantum phenomena in systems with regular geometry, local interactions and intermediate system sizes.

While here we focused on advancing the performance and generality of fully learned models, our approach could be easily combined with stronger physical priors to further improve the approximation accuracy. For example, it would be interesting to consider product states of GNN ansatz and projected pair states similar to \citep{nomura2020dirac}, as well as construct a symmetry-enhanced backflow wave-functions\cite{luo2019backflow}.

More broadly, the approaches to modeling and optimization of exponentially large objects discussed in this paper might prove useful to other applications of machine learning to scientific problems.
%
\bibliographystyle{abbrv}
\bibliography{references}

\appendix
\newpage
\section{Appendix}
\subsection{Motivation for modeling the logarithm of the wave-function }\label{apx:physical_priors}
Variational models for low energy eigenstates commonly use $\exp$ nonlinearity in the final step of the computations, which is often accomplished by working with the \emph{logarithm} of the wave-function throughout the VMC code. This choice is well supported by several observations:
\begin{enumerate}
    \item Typically only a small subset of all configurations have large magnitudes.
    \item Exponential form of $\psi_{i}$ avoids ill-behaved gradients.
    \item Solutions of the form $\psi = \exp(-Z)$ arise in quantum/stat-mech correspondence.
\end{enumerate}
The first observation is empirical and suggests that the model should have a mechanism to produce near zero outputs for a large fraction of an input domain. The second observation follows from the terms $\nabla_{w}\psi_{i} / \psi_{i}$ in the expression for the energy gradient (Eq. \ref{eq:energy_gradient_apx}). With exponential final activation this term is well behaved even for inputs $c_{i}$ for which $\psi_{i}$ is small. The last observation stems from the duality between $d$ dimensional quantum mechanics and statistical mechanics in $d+1$ dimensions.

\subsection{Stochastic estimations with Markov Chain Monte Carlo}\label{apx:mcmc}
We use Markov Chain Monte Carlo (MCMC) based on Metropolis-Hasting algorithm to stochastically estimate sums over exponentially-sized Hilbert space. The sampling is defined by the unnormalized probability distribution $P$ and a proposal function $g$. $P$ is given by the variational model as $|\psi_{f,w}(c_{i})|^{2}$. For $g$ we use a random spin exchange operation, that swaps two random spins of opposite orientation.

For efficiency we maintain a batch of MCMC chains, which evolves throughout optimization. We follow the rule of thumb of performing $N$ MCMC steps between sampling from the chain to reduce autocorrelation effects, where $N$ is equal to the number of degrees of freedom. We make sure that each chain is independent by explicitly providing different random seeds.

\subsection{Stochastic overlap estimation}\label{apx:additional_physical_observables}
We estimate the overlap $\mathcal{O}$ between two states $\ket{\psi}$ and $\ket{\phi}$ using MCMC (Eq. \ref{eq:overlap_apx}). This process involves sampling configurations from two independent Markov chains, one for each probability distribution.
\begin{align}\label{eq:overlap_apx}
    \mathcal{O} = \frac{\braket{\psi}{\phi} }{\sqrt{\braket{\psi}{\psi}\braket{\phi}{\phi}}} = \sqrt{\frac{\braket{\psi}{\phi}}{\braket{\psi}{\psi}}}\sqrt{\frac{\braket{\psi}{\phi}}{\braket{\phi}{\phi}}} &= \sqrt{\frac{\sum_{i}|\psi_{i}|^{2} \phi_{i}/\psi_{i}}{\sum_{i}|\psi_{i}|^{2}}} \sqrt{\frac{\sum_{i}|\phi_{i}|^{2} \psi_{i}/\phi_{i}}{\sum_{i}|\phi_{i}|^{2}}} \\ &= \sqrt{\E_{|\phi|^{2}}(\frac{\psi_{i}}{\phi_{i}})} \sqrt{\E_{|\psi|^{2}}(\frac{\phi_{i}}{\psi_{i}})}\nonumber
\end{align}

\subsection{Energy minimization and supervised wave-function optimization}\label{apx:optimization_methods}
Here we provide a concise summary of two optimization methods: (1) energy minimization; (2) imaginary-time supervised wave-function optimization; Both are iterative optimization schemes that rely on different gradient estimates. In this work we used the latter approach with hyperparameters discussed in \ref{apx:optimization}. For both methods it is important to take into account how the parameter updates affect the distribution of samples used to compute stochastic estimates of the quantities being optimized. For both optimization methods we provide derivations of relevant gradients in terms of $\alpha(c)=Re[\log(\psi(c))]$ and $\beta(c)=Im[\log(\psi(c))]$ which are magnitude and phase of the wave-function and outputs of our models. In all derivation we omit the explicit dependence on the parameters $w$ to reduce visual noise.

Energy minimization is based on direct minimization of $\bracket{E}$ (Eq.\ref{eq:energy}). This process is driven by the energy gradient with respect to variational parameters (Eq.\ref{eq:energy_gradient_apx}-\ref{eq:energy_gradient_final_apx}).
\begin{align}\label{eq:energy_gradient_apx}
    \nabla_{w}\bracket{E} = \nabla_{w}\frac{\bra{\psi}\hat{H}\ket{\psi}}{\braket{\psi}{\psi}} = \\ \frac{(\bra{\nabla_{w}\psi}\hat{H}\ket{\psi} + \bra{\psi}\hat{H}\ket{\nabla_{w}\psi})\braket{\psi}{\psi} - \bra{\psi}\hat{H}\ket{\psi}(\braket{\nabla_{w}\psi}{\psi} + \braket{\psi}{\nabla_{w}\psi})}{|\braket{\psi}{\psi}|^{2}} = \\
    \frac{(\bra{\nabla_{w}\psi}\hat{H}\ket{\psi} + \bra{\psi}\hat{H}\ket{\nabla_{w}\psi})}{|\braket{\psi}{\psi}|} - \frac{\bra{\psi}\hat{H}\ket{\psi}}{\braket{\psi}{\psi}} \frac{\bra{\nabla_{w}\psi}\hat{H}\ket{\psi} + \bra{\psi}\hat{H}\ket{\nabla_{w}\psi}}{\braket{\psi}{\psi}} = \\
    \frac{\sum_{i}|\psi_{i}|^{2}(\frac{\nabla_{w}\psi_{i}^{*}}{\psi_{i}^{*}}\frac{H_{ij}\psi_{j}}{\psi_{i}} + \frac{H_{ij}^{*}\psi_{j}^{*}}{\psi_{i}^{*}}\frac{\nabla_{w}\psi_{i}}{\psi_{i}})}{\sum_{i}|\psi_{i}|^{2}} - \bracket{E} \frac{\sum_{i}|\psi_{i}|^{2}(\frac{\nabla_{w}\psi_{i}}{\psi_{i}} + \frac{\nabla_{w}\psi_{i}^{*}}{\psi_{i}^{*}})}{\sum_{i}|\psi_{i}|^{2}} = \label{eq:energy_gradient_insert_identity_apx} \\
    2 \E_{|\psi|^{2}}[ Re(\frac{H_{ij}\psi_{j}}{\psi_{i}})\nabla_{w}\alpha + Im(\frac{H_{ij}\psi_{j}}{\psi_{i}})\nabla_{w}\beta] - 2 \E_{|\psi|^{2}}[\frac{H_{ij}\psi_{j}}{\psi_{i}}] \  \E_{|\psi|^{2}}[\nabla_{w}\alpha] \label{eq:energy_gradient_final_apx}
\end{align}
where in \ref{eq:energy_gradient_insert_identity_apx} we inserted an identity $\sum_{i}\ket{c_{i}}\bra{c_{i}}$ in each bracket.

The imaginary-time supervised wave-function optimization method interpolates between the energy gradient and stochastic reconfiguration\cite{sorella2007weak} approaches. The optimization process is iterative, where each iteration is accomplished via multiple SGD updates directed to maximize the overlap between the current state $\ket{\psi_{w}}$ and the target state $\ket{\phi_{r}}$, where $w$ and $r$ denote parameters of the states. The target state $\ket{\phi_{r}}$ is updated once at the beginning and each iteration and is kept fixed throughout. The imaginary-time implementation uses $\ket{\phi_{r}}=(\bm{1}-\hat{H})\ket{\psi_{r}}$, which corresponds to a small step along the imaginary-time evolution trajectory. In this case, the update of $\ket{\phi_{r}}$ at the end of the iteration reduces to setting $r \gets w$, which advances the $\ket{\phi_{r}}$ further along the imaginary-time path. The intermediate steps are accomplished by following the gradient of the negative logarithm overlap, which is derived in Eq. \ref{eq:overlap_grad_apx}-\ref{eq:overlap_grad_est_apx}.
\begin{align}
    \nabla_{w}-2\log\mathcal{O}=-\nabla_{w}\log\sqrt{\frac{\braket{\psi}{\phi}\braket{\phi}{\psi}}{\braket{\psi}{\psi}\braket{\phi}{\phi}}}=-\nabla_{w}\log(\frac{\braket{\psi}{\phi}\braket{\phi}{\psi}}{\braket{\psi}{\psi}\braket{\phi}{\phi}})= \label{eq:overlap_grad_apx} \\
    \frac{\braket{\nabla_{w}\psi}{\psi} + \braket{\psi}{\nabla_{w}\psi}}{\braket{\psi}{\psi}} - \frac{\braket{\nabla_{w}\psi}{\phi}\braket{\phi}{\psi} + \braket{\psi}{\phi}\braket{\phi}{\nabla_{w}\psi}}{\braket{\psi}{\phi}\braket{\phi}{\psi}} = \\
    \frac{\braket{\nabla_{w}\psi}{\psi} + \braket{\psi}{\nabla_{w}\psi}}{\braket{\psi}{\psi}} - \frac{\braket{\nabla_{w}\psi}{\phi}}{\braket{\psi}{\phi}} - \frac{\braket{\phi}{\nabla_{w}\psi}}{\braket{\phi}{\psi}}=\\
    \frac{\sum_{i}|\psi_{i}|^{2} (\frac{\nabla_{w}\psi_{i}^{*}}{\psi_{i}^{*}} + \frac{\nabla_{w}\psi_{i}}{\psi_{i}})}{\sum_{i}|\psi_{i}|^{2}} - \frac{\sum_{i}|\psi_{i}|^{2} \frac{\nabla_{w}\psi_{i}^{*}\phi_{i}}{\psi_{i}^{*}\psi_{i}}}{\sum_{i}|\psi_{i}|^{2}\frac{\phi_{i}}{\psi_{i}}} - \frac{\sum_{i}|\psi_{i}|^{2} \frac{\phi_{i}^{*}\nabla_{w}\psi_{i}}{\psi_{i}^{*}\psi_{i}}}{\sum_{i}|\psi_{i}|^{2}\frac{\phi_{i}^{*}}{\psi_{i}^{*}}}
    \label{eq:overlap_grad_est_apx} 
\end{align}
The final expression can be further simplified by introducing MCMC averages $r_{re}=Re[\E_{|\psi|^{2}}\frac{\phi_{i}}{\psi_{i}}]$; $r_{im}=Im[\E_{|\psi|^{2}}\frac{\phi_{i}}{\psi_{i}}]$; $\gamma=\E_{|\psi|^{2}}(Re[\frac{\phi_{i}}{\psi_{i}}]\nabla_{w}\alpha + Im[\frac{\phi_{i}}{\psi_{i}}]\nabla_{w}\beta)$; $\eta=\E_{|\psi|^{2}}(Im[\frac{\phi_{i}}{\psi_{i}}]\nabla_{w}\alpha - Re[\frac{\phi_{i}}{\psi_{i}}]\nabla_{w}\beta)$; $\delta=arg(r_{re} + i r_{im})$; using which Eq.\ref{eq:overlap_grad_est_apx} translates to:
\begin{equation}
    -2\nabla_{w}\log\mathcal{O}=\E_{|\psi|^{2}}[\nabla_{w}\alpha]-\frac{(\gamma\cos{\delta} + \eta\sin{\delta})}{\sqrt{r_{re}^{2} + r_{im}^{2}}}\label{eq:overlap_gradient_final_apx}
\end{equation}
\subsection{Test problem details}\label{apx:test_problems_details}
\subsubsection{Test problem clusters}\label{apx:test_problem_clusters}
We have trained our models to approximate the lowest energy eigenstates of the Heisenberg Hamiltonians on square, triangular and kagome lattices on different finite size clusters with periodic boundary conditions. In Table \ref{table:lattices_apx} we list the lattice vectors and periodic vectors for each cluster size.
\begin{table}[ht]
\centering
\begin{tabular}{|c|c|c|c|c|c|}
\hline 
\rule{0pt}{2.5ex}
\textbf{Lattice} & \textbf{N} & $\bm{\vec{a}_{1}}$ & $\bm{\vec{a}_{2}}$ & $\bm{\vec{b}_{1}}$ & $\bm{\vec{b}_{2}}$ \\ \hline \hline
Square & $36$ & $(1, 0)$ & $(0, 1)$ & $(6, 0)$ & $(0, 6)$ \\ \hline
Square & $64$ & $(1, 0)$ & $(0, 1)$ & $(8, 0)$ & $(0, 8)$ \\ \hline
Square & $100$ & $(1, 0)$ & $(0, 1)$ & $(10, 0)$ & $(0, 10)$ \\ \hline
Honeycomb & $32$ & $(\frac{\sqrt{3}}{2}, \frac{3}{2})$ & $(\frac{\sqrt{3}}{2}, -\frac{3}{2})$ & $(4\sqrt{3}, 6)$ & $(4\sqrt{3}, -6)$ \\ \hline
Honeycomb & $98$ & $(\frac{\sqrt{3}}{2}, \frac{3}{2})$ & $(\frac{\sqrt{3}}{2}, -\frac{3}{2})$ & $(\frac{7\sqrt{3}}{2}, \frac{21}{2})$ & $(\frac{7\sqrt{3}}{2}, -\frac{21}{2})$ \\ \hline
Triangular & $36$ & $(1, 0)$ & $(\frac{1}{2}$, $\frac{\sqrt{3}}{2})$ & $(6, 0)$ & $(3, 3\sqrt{3})$ \\ \hline
Triangular & $48$ & $(1, 0)$ & $(\frac{1}{2}$, $\frac{\sqrt{3}}{2})$ & $(6, 2\sqrt{3})$ & $(6, -2\sqrt{3})$ \\ \hline
Triangular & $108$ & $(1, 0)$ & $(\frac{1}{2}$, $\frac{\sqrt{3}}{2})$ & $(9, 3\sqrt{3})$ & $(9, -3\sqrt{3})$ \\ \hline
Kagome & $36$ & $(1, 0)$ & $(\frac{1}{2}$, $\frac{\sqrt{3}}{2})$ & $(6, -2\sqrt{3})$ & $(0, 4\sqrt{2})$ \\ \hline
Kagome & $48$ & $(1, 0)$ & $(\frac{1}{2}$, $\frac{\sqrt{3}}{2})$ & $(4, 0)$ & $(2, 2\sqrt{3})$ \\ \hline
Kagome & $108$ & $(1, 0)$ & $(\frac{1}{2}$, $\frac{\sqrt{3}}{2})$ & $(6, 0)$ & $(3, 3\sqrt{3})$ \\ \hline
\end{tabular}
\caption{Parameters of the finite size clusters.}\label{table:lattices_apx}
\end{table}
\subsubsection{Estimates of the ground state energies}\label{apx:ground_state_energies}
 For small system sizes we use exact values for $E_{0}$ whenever available. For larger system sizes exact energies are not known and we use extrapolated values $E_{0}^{*}$. For the \emph{kagome} lattice the current consensus\cite{yan2011spin} is that the energy per spin does not change noticeably beyond system size of $48$. For \emph{triangular} and \emph{honeycomb} lattices the finite size effects are believed to scale as $C \times N^{-1.5}$ with the system size due to proximity to the Neel/$120^{\circ}$ phases \cite{albuquerque2011phase,capriotti1999long}. We assumed this scaling for our estimates and extract the constant $C$ using exact and extrapolated thermodynamic limit values. All values of $E_{0}$, $E_{0}^{*}$ and corresponding references are shown in Table \ref{table:lowest_energy_states_apx}. In other cases the estimates are borrowed from the corresponding references.

\begin{table}[ht]
\centering
\begin{tabular}{|l|l|l|l|l|}
\hline
\textbf{Lattice} & \textbf{J2} & \textbf{System size} & \textbf{Energy per site} & \textbf{Reference} \\ \hline
\hline
Square & 0.5 & 36 & -0.503810 & \cite{schulz1996magnetic} \\ \hline
Square & 0.5 & 100 & -0.497629 & \cite{nomura2020dirac} \\ \hline
Triangular & 0 & 36 & -0.5603734 & \cite{capriotti1999long} \\ \hline
Triangular & 0 & TDL & -0.551 & \cite{iqbal2016spin} \\ \hline
Triangular & 0.125 & 36 & -0.515564 & \cite{iqbal2016spin} \\ \hline
Triangular & 0.125 & 108 & -0.5126 & \cite{iqbal2016spin} \\ \hline
Honeycomb & 0.2 & 32 & -0.460650 & \cite{albuquerque2011phase} \\ \hline
Honeycomb & 0.2 & TDL & -0.4527 & \cite{albuquerque2011phase} \\ \hline
Kagome & 0 & 36 & -0.43837653 & \cite{changlani2018macroscopically} \\ \hline
Kagome & 0 & 48 & -0.438703897 & \cite{lauchli2019s} \\ \hline
Kagome & 0 & 108 & -0.4386 & \cite{yan2011spin} \\ \hline
\end{tabular}
\caption{Estimates of the lowest energy eigenstates used for model comparison.}\label{table:lowest_energy_states_apx}
\end{table}
\subsection{Additional model details}\label{apx:model_details}

\subsubsection{Model architecture}
For all test problems we used the same architecture parameters to accentuate generality of the proposed method. Node and edge functions of the \textbf{Encoder}, \textbf{Processor} and \textbf{Decoder} are implemented as MLPs with $3$ hidden layers with ReLU activations. The node and edge embeddings have size $64$, while hidden layers width $128$. We use $6$ message passing steps with residual connections in the \textbf{Processor}. The final linear layer maps $128$-dimensional latent vector to $2(1)$ outputs for GNN(GNN-2) models.

For sublattice encodings ablation we used a smaller model with $2$ message passing steps; node and edge embeddings of size $16$; and hidden layers of length $16$. The reduction of the model capacity was done solely to make ablation easily reproducible in an interactive setting.

\subsubsection{Model variations}\label{apx:model_variations}
We studied several architectural variations of our model. The most notable difference is related to using separate networks to predict phases and magnitudes of the wave-functions, to which we referred to as GNN-2 in the main text. This difference persisted even in experiments where the total number of parameters between the two models was comparable.

Additionally we have considered different mechanisms for aggregating the distributed representation $Y$ (see Section \ref{sec:model}) into a latent vector $Z$. In our experiments excluding all nodes or excluding all edges from the representation did not significantly impact the performance.

\begin{figure}
  \centering
  \includegraphics[width=1.\linewidth]{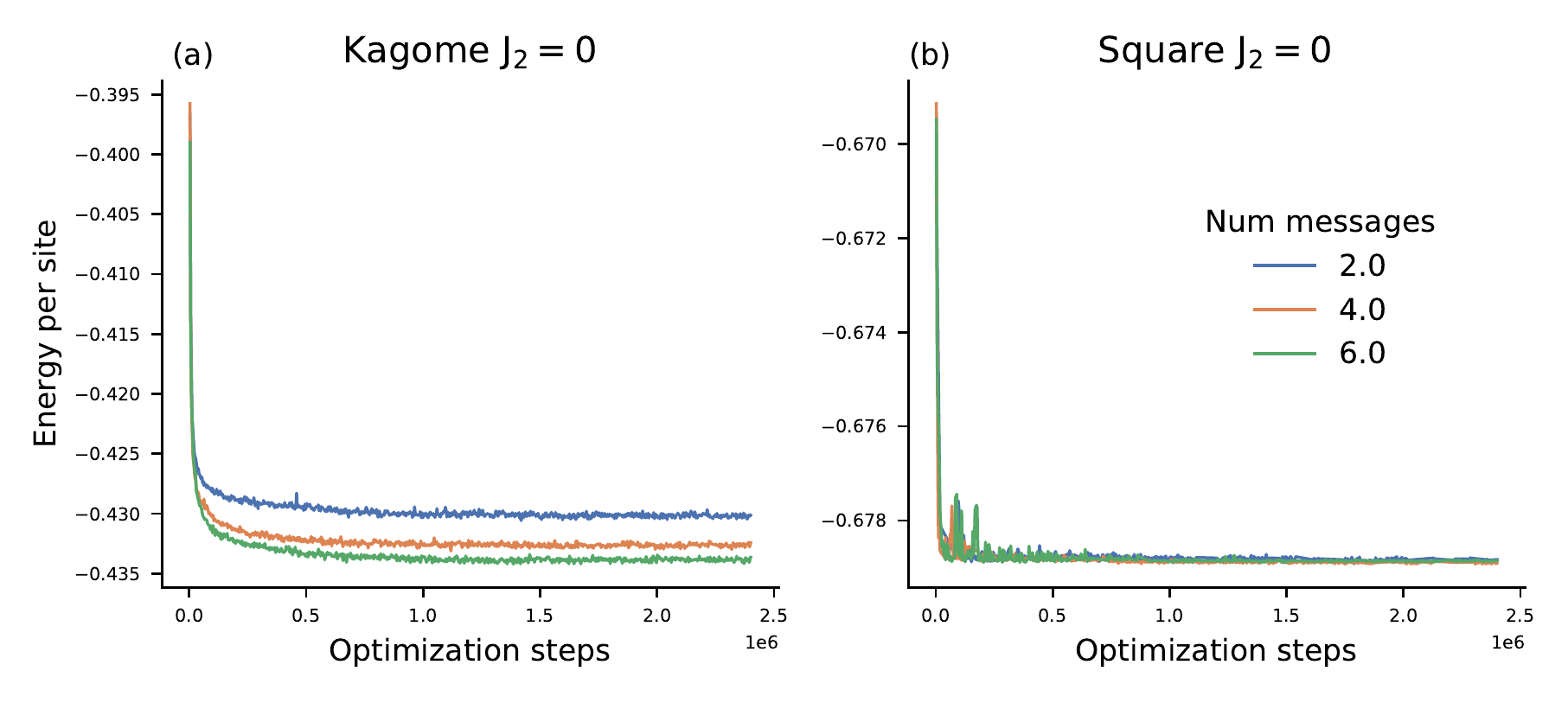}
  \caption{Varying the number of message passing steps affects the appoximation accuracy of the GNN models more when modeling disordered states. (a) Variational energy of GNN-2 model per spin as a function of optimization steps and number of message passing steps in a disordered regime on \emph{kagome} lattice and $J_{2}=0$. (b) Variational energy of GNN-2 model per spin as a function of optimization steps and number of message passing steps in an ordered regime on \emph{square} lattice and $J_{2}=0$. System size is $N=36$ for both systems.}\label{fig:num_message_passings}
\end{figure}

When varying the capacity of our models we found that in the regimes where the underlying eigenstate is ordered, even small networks perform competitively. In disordered regimes we saw a noticeable improvement in the performance as we increase the number of message passing steps, as shown in Fig.\ref{fig:num_message_passings}. Similar, albeit weaker effect was observed when the capacity of edge and node features was increased. In addition, our selective experiments show that using shared weights in the \textbf{processor} module generally achieves similar performance to the unshared variation. This suggests that more sophisticated models that select the number of message passing steps adaptively are good candidates for further improvement.

\subsubsection{Physical properties of the GNN ansatz}
From a dimensionality argument it is clear that polynomially sized variational ansatz can not represent and arbitrary quantum state. Hence it's important to understand the limitations for each class of variational wave-functions, as well as their relations to other classes.

Because by construction our ansatz represents a product of terms with local support, it can be seen as an effective implementation of a \emph{quasi-product states} introduced in  \cite{jia2020entanglement}. This suggests that the entanglement entropy of the state can obey at most an area law.

The other class of wave-functions that graph-nets ansatz is related to is correlator product states \cite{changlani2009approximating}. While any wave-function representable by our ansatz can be written as a correlator product state, it has the advantage of computational efficiency, as it works with latent representations of local states, rather than enumerating them exhaustively. This significantly decreases the number of parameters one needs to optimize.

\subsection{Sublattice encodings}\label{apx:sublattice_encodings}
\subsubsection{Symmetric sublattice encodings}\label{apx:sublattice_examples}
Our models use sublattice encoding as auxiliary input to provide a reference point for distributed computation. To enable generalization across problems of different size the encodings must respect the symmetry structure of the lattice. Our approach is inspired by regular states \cite{messio2011lattice}, in which sublattices emerge from configuration that obey lattice symmetries up to a global transformation of underlying degrees of freedom.

\begin{figure}
  \centering
  \includegraphics[width=1.\linewidth]{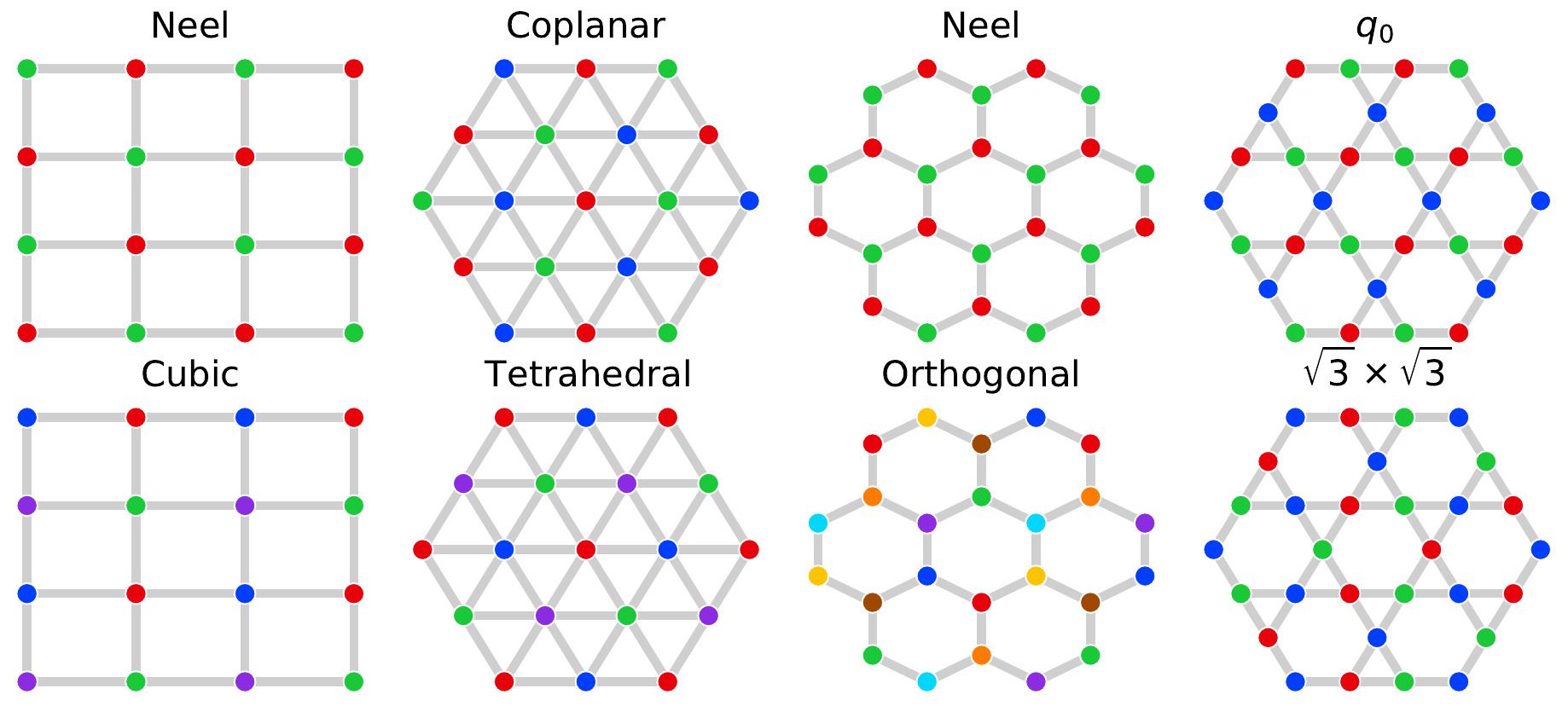}
  \caption{Sublattices used as auxiliary inputs to our model.}\label{fig:all_sublattices}
\end{figure}

All sublattices considered in this work are shown in Fig. \ref{fig:all_sublattices}. Table \ref{table:sublattices} indicates sublattices used for different values of $J_{2}$ for all $4$ lattices.
\begin{table}[ht]
\centering
\begin{tabular}{|c|c|c|c|c|c|c|c|c|}
\hline
\textbf{Lattice} & \multicolumn{2}{c|}{Square} & \multicolumn{2}{c|}{Honeycomb} & \multicolumn{2}{c|}{Triangular} & \multicolumn{2}{c|}{Kagome} \\ \hline \hline
$J_{2}$ condition & $\leq 0.5$ & $> 0.5$ & $\leq 0.5$ & $>0.5$ & $\leq0.14$ & $>0.14$ & $\geq 0$ & $< 0$ \\ \hline
Sublattice & Neel & Orth. & AF & Cubic & Coplanar & Tetr. & q=0 & $\sqrt{3}\times\sqrt{3}$ \\ \hline
\end{tabular}
\caption{Sublattice encodings used for different values of interaction strengths $J_{2}$.}\label{table:sublattices}
\end{table}

\subsubsection{Square lattice ablation results}
As stated in the main text, we found it critical to include sublattice information to obtain high quality approximations to the lowest energy eigenstates. In our ablation study, besides poor performance of a model without sublattice encodings on a $1$ dimensional spin chain, we found a different failure mode presented below.

\begin{figure}
  \centering
  \includegraphics[width=1.\linewidth]{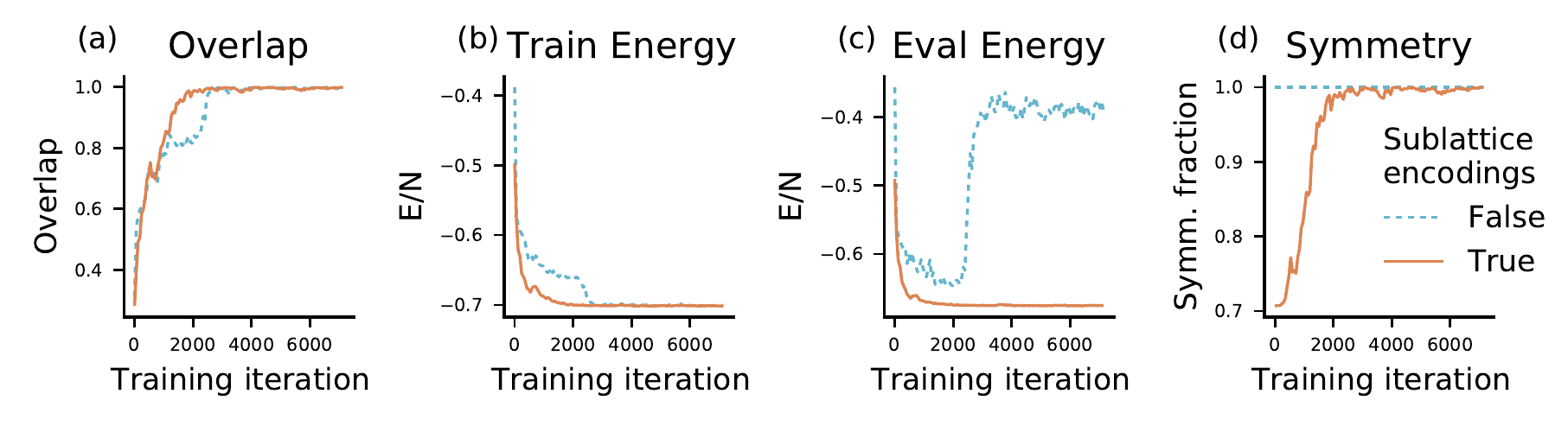}
  \caption{GNN models without sublattice encodings can learn system size dependent representations that fail to generalize to larger systems. (a) Overlap with the exact ground state on $16$ site \emph{square} lattice. (b) Energy of the models on $16$ site \emph{square} lattice. (c) Energy of the models on $36$ site \emph{square} lattice. (d) Fraction of the state in a fully symmetric subspace. Color indicates whether GNN model uses sublattice encodings. }\label{fig:sublattice_encodings_square}
\end{figure}

When the number of message passing steps of a GNN(GNN-2) model is large enough to propogate information across the entire system, the model can learn a system-size dependent approximation of the lowest energy eigenstate, which does not generalize to larger system sizes. In Fig.\ref{fig:sublattice_encodings_square} we show a similar ablation study on a square lattice. The model is optimized on a system with $16$ degrees of freedom and is additionally tested on a $36$ site system.  While the model without sublattice encodings succeeds in approximating the lowest energy eigenstate, it fails to generalize to a larger system size. The variant with sublattice encodings performs well on both system sizes.

\subsubsection{Sublattice encodings and non-universality of graph networks}\label{apx:sublattice_encodings_and_universality}

An alternative view on our use of sublattice encodings stems from non-universality of graph networks in their standard implementation \cite{maron2019universality,xu2018powerful}. By introducing a symmetry breaking encoding as an additional input to the network, we enlarge the scope of functions that can be approximated by sufficiently expressive networks. In an extreme example, if all encodings are distinct, then the corresponding graph network recovers the universal approximation property, but looses all of the spatial symmetry priors. For cases studied in this paper we found that slight symmetry breaking consistent with the underlying symmetry of the structure is usually enough to noticeably improve the performance.

\subsection{Model comparison details}

\subsubsection{Baseline models}\label{apx:baselines}
In model comparison paragraph in section \ref{sec:results} we have compared the performance of our GNN and GNN-2 models to a suite of variational baselines. Here we discuss major differences in inductive biases and neural network architectures.

Dirac spin liquid (DSL)\cite{iqbal2013gapless} and projected Bardeen-Cooper-Schrieffer (pBCS)\cite{kaneko2014gapless} ansatz are projected mean-field wave-functions. They are canonical examples of variational models that support long-range entanglement, fractionalization and topological order in spin systems\cite{iqbal2013gapless,iqbal2016spin,kolley2015phase}. These models have very few variational parameters and are interpretable, which on the other hand makes it hard to systematically improve the approximation accuracy. While our models use many more variational parameters, they systematically produce more accurate approximation to low energy states for moderate system sizes.

Neural network quantum states (CNN-C)\cite{choo2019two} is a variational ansatz based on a deep convolutional neural network with complex-valued weights. Both amplitudes $\log{|\psi_{i}|}$ and phases $arg(\psi_{i})$ are predicted simultaneously via a single network pass, where a generalized ReLU\cite{trabelsi2017deep} is used as an activation function. To stabilize the process of learning the phase structure of the low energy states this model additionally includes a reference sign structure given by the Marshal sign rule\cite{marshall1955antiferromagnetism}. Our models differ from this architecture in several ways: (1) GNN-2 models amplitudes and phases separately; (2) both GNN models are applicable to arbitrary lattices and incorporate lattice symmetries beyond translation.

In \cite{szabo2020neural} (CNN-P), authors propose to model amplitudes $\log{|\psi_{i}|}$ and phases $arg(\psi_{i})$ separately, similar to our GNN-2 model. Two main differences between GNN-2 and CNN-P include: (1) applicability of GNN based models to arbitrary lattices; (2) parameterization of the phase prediction. In CNN-P the phase is computed by summing locally predicted phasors of unit length and computing the global angle. This is equivalent to sum-based reduction over $\sin$ and $cos$ components of the predicted phases. GNN-2 uses sum-based reduction directly over the angles, which makes it possible for a local change to noticeably affect the phase of the wave-function, which is known to be highly oscillatory.

Neural Gutzwiller-projected states (GW-RBM)\cite{ferrari2019neural} and neural net backflow (NN-BF)\cite{luo2019backflow} are examples of hybrid approaches that use ML to build on-top of existing variational models. GW-RBM extends the projected mean-field approach by weighting the contributions of orbitals, with coefficients predicted by a RBM architecture. NN-BF draws on the ideas of backflow\cite{feynman1956energy}, where the single particle orbitals are made configuration dependent. GNN models considered in this paper do not include such inductive biases and hence do not rely on existence of a reference ansatz.

\subsubsection{Model performance}\label{apx:model_performance}
We provide numeric values of variational energies for our GNN, GNN-2 models and ResNet baseline in Table \ref{table:energies}. All results were computed using MCMC using $2 * 10^{5}$ samples from equilibrated Markov chains. The stochastic estimation error bars are marked by parenthesis around the last significant digit.

\begin{table}[ht]
\centering
\begin{tabular}{|c|c|c|c|c|c|c|c|}
\hline
\textbf{Lattice} & \textbf{Parameters} & \multicolumn{2}{c|}{GNN} & \multicolumn{2}{c|}{GNN-2} & \multicolumn{2}{c|}{ResNet} \\ \hline
\multirow{3}{*}{Square} & $J_{2}$ & 0 & 0 & 0 & 0 & 0 & 0 \\ \cline{2-8} 
 & $N$ & \multicolumn{2}{c|}{100} & \multicolumn{2}{c|}{100} & \multicolumn{2}{c|}{100} \\ \cline{2-8} 
 & $\langle E \rangle / N$ & \multicolumn{2}{c|}{-0.6708(0)} & \multicolumn{2}{c|}{-0.5023(5)} & \multicolumn{2}{c|}{-0.6705(2)} \\ \hline
\multirow{3}{*}{Square} & $J_{2}$ & 0.5 & 0.5 & 0.5 & 0.5 & 0.5 & 0.5 \\ \cline{2-8} 
 & $N$ & 36 & 100 & 36 & 100 & 36 & 100 \\ \cline{2-8} 
 & $\langle E \rangle / N$ & -0.5022(4) & -0.4955(4) & -0.5023(5) & -0.4960(5) & -0.5017(7) & -0.488(7) \\ \hline
\multirow{3}{*}{Honeycomb} & $J_{2}$ & 0.2 & 0.2 & 0.2 & 0.2 & 0.2 & 0.2 \\ \cline{2-8} 
 & $N$ & 32 & 98 & 32 & 98 & 32 & 98 \\ \cline{2-8} 
 & $\langle E \rangle / N$ & -0.4563(6) & -0.4528(2) & -0.4564(7) & -0.4536(5) & -0.4561(1) & -0.4496(4) \\ \hline
\multirow{3}{*}{Triangular} & $J_{2}$ & 0 & 0 & 0 & 0 & 0 & 0 \\ \cline{2-8} 
 & $N$ & \multicolumn{2}{c|}{108} & \multicolumn{2}{c|}{108} & \multicolumn{2}{c|}{108} \\ \cline{2-8} 
 & $\langle E \rangle / N$ & \multicolumn{2}{c|}{-0.5508(8)} & \multicolumn{2}{c|}{-0.5519(4)} & \multicolumn{2}{c|}{-0.5451(2)} \\ \hline
\multirow{3}{*}{Triangular} & $J_{2}$ & 0.125 & 0.125 & 0.125 & 0.125 & 0.125 & 0.125 \\ \cline{2-8} 
 & $N$ & 36 & 108 & 36 & 108 & 36 & 108 \\ \cline{2-8} 
 & $\langle E \rangle / N$ & -0.512(2) & -0.500(9) & -0.5131(8) & -0.5069(8) & -0.5007(5) & -0.488(9) \\ \hline
\multirow{3}{*}{Kagome} & $J_{2}$ & 0 & 0 & 0 & 0 & 0 & 0 \\ \cline{2-8} 
 & $N$ & 36 & 108 & 36 & 108 & 36 & 108 \\ \cline{2-8} 
 & $\langle E \rangle / N$ & -0.434(1) & -0.4302(1) & -0.4338(6) & -0.4314(7) & -0.433(0) & -0.424(9) \\ \hline
\end{tabular}
\caption{Numerical values of variational energies per spin for GNN, GNN-2 models and ResNet baseline.}\label{table:energies}
\end{table}

\subsubsection{Computational efficiency}\label{apx:computational_efficiency}
Computational complexity of variational approach is largely determined by the efficiency of evaluating the wave-function amplitudes for a batch of configurations. To measure computational efficiency we have timed the process of evaluating $7200$ batches of $32$ basis configuration within a MCMC process. The GNN-2 model took $17.4$ seconds to accomplish this task, while our unstructured ResNet baseline finishes in $2.14$ seconds. While GNN-based models are $4-8\times$ slower, they outperform the unstructured models in approximation quality (even when the number of training iterations of the unstructured model is increased appropriately).

\subsection{Optimization parameters}\label{apx:optimization}
We optimized all models using IT-SWO\cite{kochkov2018variational}. We used $30$ iterations in the inner loop and chose imaginary time constant $\beta$ based on the relative scale of the energy spectrum for different problems. The exact values are listed in Table \ref{table:ite_beta_apx}. Each iteration was performed using adam optimizer\cite{kingma2014adam} with $b_{1}=0.9$; $b_{2}=0.99$ and learning rate that was decayed exponentially with the rate of $0.1$ over $8*10^{5}$ updates starting with $7*10^{-4}$ ($4*10^{-4}$ for ResNet baseline, which resulted in better performance). During training we use the same MCMC parameters as for evaluation discussed in Appendix \ref{apx:mcmc}.

We optimized all GNN models for $8*10^{5}$ iterations ($\approx 26000$ IT-SWO steps). For larger system sizes we initialized the weights with values from smaller system sizes. When training a partial ResNet baseline we used same parameters but trained it for $2*10^{6}$ steps.

All models were optimized using Google's Cloud TPU v4. All models trained within a day.

\begin{table}[ht]
\centering
\begin{tabular}{|l|c|c|c|c|c|c|c|c|}
\hline
 \textbf{System} & \multicolumn{2}{c|}{Square} & \multicolumn{2}{c|}{Honeycomb} & \multicolumn{2}{c|}{Triangular} & \multicolumn{2}{c|}{Kagome} \\ \hline \hline
N & 36 & 100 & 32 & 98 & 36 & 108 & 36 & 108 \\ \hline
$\bm{\beta}$ & 0.04 & 0.007 & 0.04 & 0.01 & 0.04 & 0.01 & 0.05 & 0.015 \\ \hline
\end{tabular}
\caption{Parameters of the imaginary time evolution step parameter used for optimization on different systems.}\label{table:ite_beta_apx}
\end{table}
\end{document}